\documentclass[12pt]{article}
\usepackage{epsfig,color}

\usepackage{graphicx,epsf,amssymb}

\newcommand\pubnumber{SLAC--PUB--12034}
\newcommand\pubdate{\today}
\newcommand\hepnumber{hep-ph/0608027}

\def\SLAC{Stanford Linear Accelerator Center\\
    Stanford University, Stanford, California 94309 USA}
\def\doeack{\footnote{Work supported by the Department of Energy,
                     contract DE--AC02--76SF00515.}}

\def\Title#1{\begin{center} {\Large #1 } \end{center}}
\def\Author#1{\begin{center}{ \sc #1} \end{center}}
\def\Address#1{\begin{center}{ \it #1} \end{center}}

\newcommand\pubblock{\rightline{\begin{tabular}{l} \pubnumber\\
         \pubdate \\ \hepnumber \end{tabular}}}
\newenvironment{Abstract}{\begin{quotation} \begin{center}
                       ABSTRACT
     \end{center}\bigskip  }{\end{quotation}}

\def\Acknowledgements{\bigskip  \bigskip \begin{center} \begin{large}
             \bf Acknowledgements \end{large}\end{center}}

\def\fig#1{fig.~{\ref{#1}}}

\def \be  {\begin{equation}}
\def \ee  {\end{equation}}
\def \bea  {\begin{eqnarray}}
\def \eea  {\end{eqnarray}}
\def\eqn#1{eq.~(\ref{#1})}
\def\Eqn#1{Eq.~(\ref{#1})}

\newcommand{\nn}{\nonumber}

\def\Li{\mathop{\rm Li}\nolimits}
\def\Ls{\mathop{\rm Ls}\nolimits}
\def\Ll{\mathop{\rm L{}}\nolimits}
\def\Res{\mathop{\rm Res}}

\def\Ksl{\s{K}}

\def\NeqFour{{\cal N}=4}
\def\NeqOne{{\cal N}=1}
\def\NeqZero{{\cal N}=0}
\def\tlambda{{\tilde\lambda}}
\def\eps{\epsilon}

\def\nf{n_{\mskip-2mu f}}
\def\Nc{N_{c}}
\def\Shift#1#2{{[#1,#2\rangle}}
\def\cg{c_\Gamma}
\def\Vertex{R}

\def\DiagrammaticRational{R^D}

\def\PureCut{C}
\def\Overlap{O}
\def\CuthRat{{\widehat {CR}}}
\def\Remaining{{\widehat {R}}}
\def\Cuth{{\widehat {C}}}
\def\Inf{\mathop{\rm Inf}}

\def\tree{{\rm tree}}
\def\nonst{{\rm \, non-standard}}
\def\rec{{\rm \, recursive}}
\def\Fact{{\cal F}}
\def\DiagrammaticRationalSrec#1{R^{D,\rec\,#1}}
\def\DiagrammaticRationalSnonst#1{R^{D,\nonst\,#1}}
\def\InfPart#1#2{\mathop{\rm Inf}\limits_{#1}{#2}}

\def\Lzz{\mathop{\hbox{\rm L}}\nolimits_2}
\def\Lz{\mathop{\hbox{\rm L}}\nolimits_0}
\def\Kz{\mathop{\hbox{\rm K}}\nolimits_0}

\def\Kh{{\hat K}}

\def\spa#1.#2{\left\langle#1\,#2\right\rangle}
\def\spb#1.#2{\left[#1\,#2\right]}
\def\spab#1.#2.#3{\sandmm#1.#2.#3}
\def\spba#1.#2.#3{\sandpp#1.#2.#3}
\def\spaa#1.#2.#3.#4{\sandmp#1.{#2#3}.#4}
\def\spbb#1.#2.#3.#4{\sandpm#1.{#2#3}.#4}
\def\spash#1.#2{\vphantom{\hat K}\spa{\smash{#1}}.{\smash{#2}}}
\def\spbsh#1.#2{\vphantom{\hat K}\spb{\smash{#1}}.{\smash{#2}}}
\def\lor#1.#2{\left(#1\,#2\right)}
\def\sand#1.#2.#3{%
\left\langle\smash{#1}{\vphantom1}^{-}\right|{#2}%
\left|\smash{#3}{\vphantom1}^{-}\right\rangle}
\def\sandp#1.#2.#3{%
\left\langle\smash{#1}{\vphantom1}^{-}\right|{#2}%
\left|\smash{#3}{\vphantom1}^{+}\right\rangle}
\def\sandpp#1.#2.#3{%
\left\langle\smash{#1}{\vphantom1}^{+}\right|{#2}%
\left|\smash{#3}{\vphantom1}^{+}\right\rangle}
\def\sandpm#1.#2.#3{%
\left\langle\smash{#1}{\vphantom1}^{+}\right|{#2}%
\left|\smash{#3}{\vphantom1}^{-}\right\rangle}
\def\sandmp#1.#2.#3{%
\left\langle\smash{#1}{\vphantom1}^{-}\right|{#2}%
\left|\smash{#3}{\vphantom1}^{+}\right\rangle}
\def\sandmm#1.#2.#3{%
\left\langle\smash{#1}{\vphantom1}^{-}\right|{#2}%
\left|\smash{#3}{\vphantom1}^{-}\right\rangle}
\def\spab#1.#2.#3{\sandmm#1.#2.#3}
\def\spbb#1.#2.#3.#4{\sandpm#1.{#2#3}.#4}
\newbox\charbox
\newbox\slabox
\def\s#1{{      
        \setbox\charbox=\hbox{$#1$}
        \setbox\slabox=\hbox{$/$}
        \dimen\charbox=\ht\slabox
        \advance\dimen\charbox by -\dp\slabox
        \advance\dimen\charbox by -\ht\charbox
        \advance\dimen\charbox by \dp\charbox
        \divide\dimen\charbox by 2
        \raise-\dimen\charbox\hbox to \wd\charbox{\hss/\hss}
        \llap{$#1$}
}}

\def \as {\relax\ifmmode\alpha_s\else{$\alpha_s${ }}\fi}

\begin{document}
\begin{titlepage}
\pubblock

\vfill
\Title{\bf{Bootstrapping One-Loop QCD Amplitudes}
\footnote{Work in collaboration with Zvi Bern, Lance Dixon, Darren Forde,
and David Kosower.}~\footnote{
 Extended version of talks given \\
at the \textit{7th Workshop On
Continuous Advances In QCD},
11-14 May 2006, Minneapolis, Minnesota; \\
at \textit{SUSY06: 14th International Conference On Supersymmetry
And The Unification Of Fundamental Interactions},
12-17 Jun 2006, Irvine, California;\\
at the
\textit{LoopFest V: Radiative Corrections For The International
Linear Collider: Multi-Loops And Multi-Legs},
19-21 Jun 2006, SLAC, Menlo Park, California;\\
and at the \textit{Vancouver Linear Colliders Workshop (ALCPG 2006)},
19-22 Jul 2006, Vancouver, British Columbia.}
}
\vfill

\Author{ Carola F. Berger\doeack}
\Address{\SLAC}
\vfill
\begin{Abstract}
We review the recently developed bootstrap method for the computation
of high-multiplicity QCD amplitudes at one loop.
We illustrate the general algorithm step by step with a six-point example.
The method combines
(generalized) unitarity with on-shell recursion relations to determine
the not cut-constructible, rational terms of these amplitudes. Our bootstrap
approach works for arbitrary configurations of gluon
helicities and arbitrary numbers of external legs.
\end{Abstract}
\vfill

\end{titlepage}
\def\thefootnote{\fnsymbol{footnote}}
\setcounter{footnote}{0}

\section{Introduction}

The Large Hadron Collider (LHC), which is scheduled to begin operation in
2007, will provide new insight into the main missing piece of the Standard
Model, the origin of electroweak symmetry breaking, and potentially discover new physics
beyond the Standard Model. Discoveries at the LHC require a thorough
understanding of Standard Model processes, all of which have many
particles in the final state. Similarly, very precise knowledge of
backgrounds and expected signals at the future International Linear
Collider is required. In order to fully exploit the potential of both
machines, many high-multiplicity processes need to be computed to as high
accuracy as possible, which entails the computation of multi-leg amplitudes
to at least one-loop order. This is a challenging endeavor,  and
traditional methods based on the direct analytical evaluation of an enormous
magnitude of Feynman diagrams are not optimized for this task.

Here we review a recently developed method that combines on-shell recursion
relations with (generalized) unitarity to compute one-loop multi-gluon
amplitudes recursively, thereby ``recycling'' information from amplitudes
with fewer legs \cite{Bern:2005hs,Berger:2006ci,Berger:2006vq}.
The bootstrap method works for arbitrary helicity configurations. Moreover,
in some cases, the recursion can be solved explicitly, yielding all-multiplicity
expressions for one-loop gluon amplitudes
\cite{Bern:2005hs,Berger:2006ci,Berger:2006vq,Forde:2005hh,FordeProcs}.
Our bootstrap method
relies only on factorization properties of the amplitudes as well as
unitarity, and should therefore be amenable to an extension to amplitudes
with external fermions and massive partons.

After a brief summary of our notation,
we review the on-shell recursion relations for tree-level
amplitudes derived via Cauchy's theorem by continuing the physical
amplitudes into the complex plane. At one loop, several new features arise,
and Cauchy's theorem is not immediately applicable.
We show how to overcome these difficulties with the bootstrap approach,
which combines the information from unitarity and two continuations
in independent complex parameters. We summarize our results and
some open problems and give an outlook on possible future directions and
applications of the method.

\section{Notation}

In the following, we use the spinor helicity formalism (see for example
\cite{Dixon:1996wi} and references therein) to express the amplitudes in terms of
spinor inner products,
\be
\spa{j}.{l} = \langle j^- | l^+ \rangle = \bar{u}_-(k_j) u_+(k_l)\,,
\hskip 2 cm
\spb{j}.{l} = \langle j^+ | l^- \rangle = \bar{u}_+(k_j) u_-(k_l)\, ,
\label{spinorproddef}
\ee
where $u_\pm(k)$ is a massless Weyl spinor with momentum $k$ and positive
or negative chirality, respectively, which we also write as,
\begin{equation}
 \lambda_i \equiv u_+(k_i), \qquad \tlambda_i \equiv u_-(k_i) \,.
\label{lambdadef}
\end{equation}
Furthermore, we strip off all color information, as well as coupling
constants, and compute only the leading-color amplitudes.
We employ the following supersymmetric decomposition of the leading-color
one-loop QCD amplitudes in the four-dimensional helicity
scheme~\cite{Bern:1993mq},
\begin{eqnarray}
A_{n;1}^{\rm QCD} &=&
A^{\NeqFour}_{n;1} -4 A^{\NeqOne}_{n;1}+  A^{\NeqZero}_{n;1}
+ {\nf\over\Nc} \Bigl( A^{\NeqOne}_{n;1}- A^{\NeqZero}_{n;1}\Bigr) \,,
\label{AnQCD}
\end{eqnarray}
where $\nf$ is the number of active quark flavors in QCD.
The contributions from the $\NeqFour$ and $\NeqOne$ supersymmetric multiplets
are completely cut-constructible, that is, they contain only (poly)logarithms
and associated constants ($\pi^2$) and are thus computable via (generalized)
unitarity. The remaining non-supersymmetric $\NeqZero$ contribution
from a scalar running in the loop has both cut-constructible and
rational parts. It is the latter part that we will compute via on-shell
recursion relations. In the following we will suppress the $\NeqZero$ superscript
and write $A_n \equiv A^{\NeqZero}_{n;1}$.

For further information and our notational
conventions we refer to \cite{Berger:2006ci} and the lecture notes
\cite{Dixon:1996wi}.

\section{On-Shell Recursion Relations at Tree Level}

\subsection{Review of the Proof at Tree Level}

Here we briefly review the on-shell recursion
relations found in ref.~\cite{Britto:2004ap} and proven in
ref.~\cite{Britto:2005fq} for tree level amplitudes.
For further details we refer to these papers.

At tree level, the on-shell recursion relations rely on general properties
of complex functions as well as factorization properties of
scattering amplitudes.
The proof~\cite{Britto:2005fq} of the tree-level relations employs a
parameter-dependent ``$[j,l\rangle$''
shift of two of the external massless spinors, $j$ and $l$,
in an $n$-point process,
\begin{equation}
\Shift{j}{l}:\hskip 2 cm
\tlambda_j \rightarrow \tlambda_j - z\tlambda_l \,,
\hskip 2 cm
\lambda_l \rightarrow \lambda_l + z\lambda_j \,.
\label{SpinorShift}
\end{equation}
where $z$ is a complex number.  The corresponding momenta
are then continued in the complex plane as well
so that they remain massless, $k_j^2(z) = 0 = k_l^2(z)$,
and overall momentum conservation is maintained.

An on-shell amplitude containing the momenta $k_j$ and $k_l$
then also becomes parameter-dependent,
\begin{equation}
A(z) = A(k_1,\ldots,k_j(z),k_{j+1},\ldots,k_l(z),\ldots,k_n).
\end{equation}
When $A$ is a tree amplitude or finite one-loop
amplitude, $A(z)$ is a rational function of $z$.
The physical amplitude is given by $A(0)$.

We can then use Cauchy's theorem,
\begin{equation}
{1\over 2\pi i} \oint_C {dz\over z}\,A(z)  = 0\,,
\label{ContourInt}
\end{equation}
where the contour $C$ is taken around the circle at infinity,
and the integral vanishes if the
complex continued amplitude $A(z)$ vanishes as $z \rightarrow \infty$.
Evaluating the integral as a sum
of residues, we can then solve for the physical amplitude $A(0)$ to obtain,
\begin{equation}
A(0) = -\sum_{{\rm poles}\ \alpha} \Res_{z=z_\alpha}  {A(z)\over z}\,.
\label{NoSurface}
\end{equation}

\begin{figure}[t]
\centerline{\epsfxsize 2 truein\epsfbox{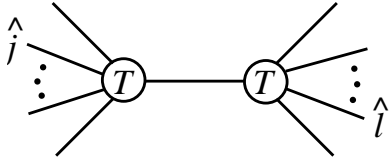}}
\caption{Schematic representation of a typical tree recursive
contribution to \eqn{BCFW}.  The labels `$T$' refer to
on-shell tree amplitudes. The momenta $\hat{j}$ and $\hat{l}$
are complex continued, on-shell momenta. }
\label{TreeGenericFigure}
\end{figure}

\begin{figure}[bht]
\begin{center}
\hspace*{-2.3cm}\parbox{2.5cm}{\epsfysize=38mm \epsfbox{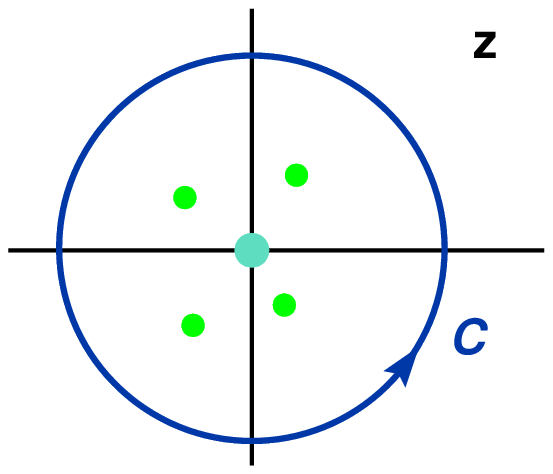}}
\hspace*{3cm}
\parbox{2cm}{\epsfysize=40mm \epsfbox{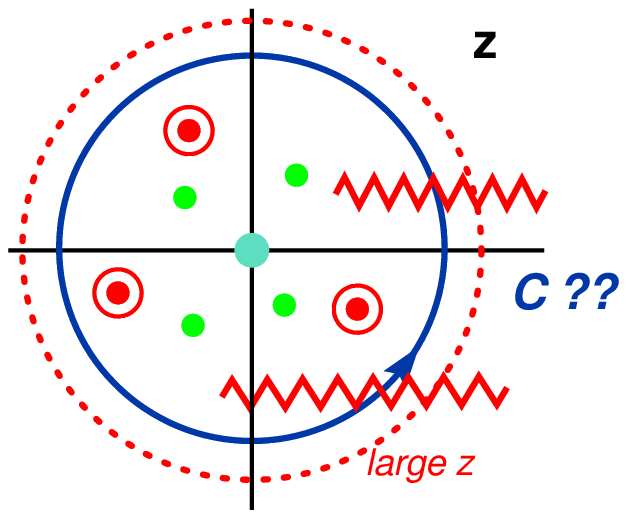}}
\end{center}
\centerline{ \hspace*{3cm} (a) \hspace*{4.8cm} (b) \hspace*{3.6cm}}
\caption{
Illustration of the pole-structure in the complex plane for QCD amplitudes at
(a) tree level, and (b) at the one-loop level. The point at the origin
symbolizes
the physical amplitude at $z = 0$. Branch cuts are drawn as zigzagged lines.
Simple poles are depicted as green dots,
non-standard poles as red circled dots. The dashed line symbolizes the
large-parameter contribution.
}
\label{Cartoon}
\end{figure}

As explained in ref.~\cite{Britto:2005fq}, at tree level $A(z)$ only has simple poles,
as schematically illustrated in fig.~\ref{Cartoon}~(a). These poles arise when
 shifting the propagators in the amplitude. For example,
\be
\frac{i}{K_{r \dots l \dots s}^2} \rightarrow  \frac{i}{K_{r \dots s}^2 + z
\sand{j}.{\Ksl_{r\ldots s}}.{l} } \, , \label{propagator}
\ee
if the set of legs $\{r, \dots ,s\}$ contains leg $l$,
which is shifted according
to \eqn{SpinorShift}. The complex continued
amplitude is then schematically given by,
\be
A \rightarrow A(z) = \sum_{r,s}
\sum_{h} A^h_{L}(z) {i \over
K_{r \dots s}^2 + z
\sand{j}.{\Ksl_{r\ldots s}}.{l}  } A^{-h}_R(z) \, ,
\ee
where $h=\pm1$ labels the helicity of the intermediate state, and
the labels $L$ and $R$ denote amplitudes with fewer legs, which
the propagator \eqn{propagator} connects.
There is generically a double sum, labelled by $r,s$,
dividing the set of legs into partitions
with legs $j$ and $l$ always appearing on opposite sides of the $z$-dependent
propagator.
Therefore, each residue in \eqn{NoSurface}
is given by factorizing the shifted amplitude on the
 poles in momentum invariants, so that at tree level,
\begin{equation}
A(0) = \sum_{r,s} \sum_{h}
    A^h_{L}(z = z_{rs}) { i \over K_{r\cdots s}^2 } A^{-h}_R(z = z_
{rs})  \,.
\label{BCFW}
\end{equation}
The squared momentum associated with that pole,
$K_{r\cdots s}^2$, is evaluated in the unshifted kinematics;
whereas the on-shell amplitudes with fewer legs, $A_{L}$ and $A_R$,
are evaluated in kinematics that have been shifted by \eqn{SpinorShift} with
$z=z_{rs}$, where \eqn{propagator} has a pole,
\begin{equation}
z_{rs} = - {K_{r\cdots s}^2 \over \sand{j}.{\Ksl_{r\ldots s}}.{l} } \,.
\end{equation}
In the following, such shifted, on-shell momenta will be denoted by
$k(z = z_{rs}) \equiv \hat{k}$.
A typical contribution to the sums in \eqn{BCFW} is
illustrated in fig.~\ref{TreeGenericFigure}.

We have thus succeeded in expressing the $n$-point amplitude $A$ in terms
of sums over on-shell, but complex continued, amplitudes with fewer legs,
which are connected by scalar propagators. For certain helicity
configurations, this recursion relation can be solved explicitly, leading
to new all-multiplicity expressions for these amplitudes.

\subsection{New Features at the One-Loop Level}

The proof reviewed above relies on two properties of the complex continued
tree level amplitude:
\begin{itemize}
\item The amplitude only has simple poles, which correspond to
\emph{physical} multiparticle
and collinear factorizations.
\item The amplitude vanishes at infinity.
\end{itemize}
These properties do not depend on the specific details of the gauge
theory under
consideration. Therefore, at tree level, the application of the
recursion relation \eqn{BCFW}
has been extended beyond multi-gluon amplitudes,
to the computation of amplitudes including fermions, scalars, and
 massive partons, to the computation
of gravity amplitudes, and of Higgs amplitudes produced via
a heavy quark loop, in the effective theory where the heavy quark
has been integrated out.
Further details and references can be found in 
refs.~\cite{FordeProcs,Cachazo:2005ga}.

At loop level, however, these
properties do not hold in general, and several new features arise:
\begin{itemize}
\item The (poly)logarithmic terms of the amplitude have spurious,
unphysical, singularities, which however cancel in the full amplitude
against corresponding terms in the rational part. These
spurious singularities are already present in real kinematics.
For example, terms of the following form arise,
\begin{equation}
{\ln(r)\over (1-r)^2} \,,
\label{sampleL1pure}
\end{equation}
which is singular as $r \rightarrow 1$. Such spurious singularities
complicate the application
of Cauchy's theorem because it requires to sum over all poles,
whether physical or unphysical.
\item The amplitude develops branch cuts in the complex plane
that stem from the complex continuation of (poly)logarithms.
\item The complex continued amplitude contains double poles
$\sim \frac{ \spa{a}.{b}}{{\spb{a}.{b}}^2}$, and
`unreal' poles $\sim \frac{ \spa{a}.{b}}{{\spb{a}.{b}}}$.
In real kinematics, expressions of this form have
only a single pole, or are not singular, respectively,
because the spinor products $\spa{a}.{b}$ and $\spb{a}.{b}$
differ only by a phase. With the complex continuation \eqn{SpinorShift}
these ratios of spinor products develop (double) poles.
Such poles arise in two-particle channels with like helicities,
and the complex factorization properties in these
channels are not yet fully understood.
We term these channels `non-standard'.
\item The amplitude does not vanish as $z\rightarrow \infty$.
\end{itemize}
These new features are illustrated in  fig.~\ref{Cartoon}~(b).
For general helicity configurations, as explained in ref.~\cite{Berger:2006ci},
one can choose a pair of shifted momenta that avoids either non-standard
channels or large-parameter contributions, but in general not both.

\section{The Bootstrap Method}

The bootstrap method developed in refs.~\cite{Bern:2005hs,Berger:2006ci}
systematically deals with the aforementioned complications. We use the
term
`bootstrap' because we
``assume that very general consistency criteria are
sufficient to determine the whole theory completely''~\cite{wiki}.
In the present case, the consistency criteria we employ
are unitarity, to determine the branch cuts, and factorization,
to obtain the rational remainder via on-shell recursion relations.
The self-consistency of the approach and thus the
correctness of the results is checked via collinear and multiparticle
factorization limits in all possible channels.
Any omitted terms would spoil the correct factorization behavior.

\subsection{Branch Cuts from Unitarity}

In order to deal with branch cuts and associated spurious singularities,
we begin by decomposing the amplitude into
`pure-cut' and `rational' parts,
\begin{equation}
A_n(z) =  \cg \Bigl[ \PureCut_n(z) + \Vertex_n(z) \Bigr] \,,
\label{ACREq}
\end{equation}
where we have explicitly taken an ubiquitous one-loop factor
$\cg$ outside of $\PureCut_n(z)$ and $\Vertex_n(z)$,
\begin{equation}
\cg = {1\over(4\pi)^{2-\eps}}
  {\Gamma(1+\eps)\Gamma^2(1-\eps)\over\Gamma(1-2\eps)}\, .
\label{cgdefn}
\end{equation}
The rational parts are defined by setting all logarithms, polylogarithms,
and associated $\pi^2$ terms to zero,
\begin{equation}
\Vertex_n(z) \equiv {1\over \cg} A_n(z)\Bigr|_{\rm rat} =
{1\over \cg} A_n(z)\biggr|_{\ln, \Li, \pi^2 \rightarrow 0} \,.
\label{RationalDefinition}
\end{equation}
The pure-cut terms are the remaining terms, all of which must
contain logarithms, polylogarithms, or $\pi^2$ terms.
In the following
we assume that the cut containing terms have been computed via
(generalized) unitarity or other means (see, e.\,g.
\cite{Bedford:2004nh,Bern:2005hh,Britto:2006sj}
and references therein), and derive an
on-shell recursion for the rational remainder.

As mentioned above, the presence of spurious, unphysical singularities
in the pure-cut parts complicates
this task. These spurious poles cancel in the full amplitude.
We will therefore `complete the cut' by adding
rational terms $\CuthRat$ to the pure-cut parts that eliminate
these spurious poles
from the beginning.
For example, we add a term proportional to $\sim \frac{1}{1-r}$
to \eqn{sampleL1pure} to eliminate the singularity as $r \rightarrow 1$.
Effectively, we replace certain combinations of (poly)logarithms
by $\Ll_i$- and $\Ls_i$-functions (see e.\,g.
ref.~\cite{Berger:2006vq} for definitions). For
example, we replace \eqn{sampleL1pure} with
\be
\Ll_1(r) = {\frac{\ln(r)}{(1-r)}+1\over 1-r}\, .
\ee
Instead of the decomposition \eqn{ACREq} we now have,
\begin{equation}
A_n(z) = \cg \Bigl[
\bigg( C_n(z) + \CuthRat_n(z) \bigg) +
\bigg( R_n(z) - \CuthRat_n(z) \bigg) \Bigr]
\equiv \cg \Bigl[ \Cuth_n(z) + \Remaining_n(z) \Bigr] \, ,
\label{CompletedCutDecomposition}
\end{equation}
where we have defined $\Cuth \equiv C + \CuthRat$ and
$\Remaining = R - \CuthRat$.
Of course, such a cut-completion is not unique. Also, when
constructing a recursion relation for the rational remainder, we need
to take into account that $\Cuth_n$ already contains rational terms
from this cut-completion in order to avoid double counting.
The resulting full amplitude is then unambiguous.

\subsection{On-Shell Recursion for Rational Terms}

Since cut and rational parts factorize separately~\cite{Bern:2005hs}, we
can now apply Cauchy's theorem to \eqn{CompletedCutDecomposition}.
Note that $\Cuth_n(z)$ may have (spurious) contributions as $z
\rightarrow \infty$, denoted by
$\Inf \Cuth_n$, which we can subtract off, since we know
$\Cuth_n(0)$ (and thus $\Cuth_n(z)$). We obtain,
\be
A_n(0) = \Inf A_n + \cg \Biggl[ \Cuth_n(0)  - \Inf \Cuth_n
-\sum_{{\rm poles}\ \alpha} \Res_{z=z_\alpha}  {R_n(z)\over z}
+ \sum_{{\rm poles}\ \alpha} \Res_{z=z_\alpha}  {\CuthRat_n(z)\over z} \Biggr]\,.
\label{loopcontour}
\ee
Here, we have taken into account the fact that the full amplitude does not
necessarily vanish at infinity under the chosen shift. This contribution
is denoted by $\Inf A_n$. Its formal operator definition is the
extraction of the constant $z^0$ term
in a Laurent expansion of $A_n(z)$ around $z = \infty$,
\be
\Inf A_n(z) = \Inf \, \left[ \sum_i \left( a^{(i)} z^i
+ \frac{b^{(i)} + \ln(1/z) c^{(i)} + \ln^2(1/z) d^{(i)}}{z^i} \right)
\right] = a^{(0)} \, . \label{Infdef}
\ee
The subtraction of $\Inf \Cuth_n$ is necessary, because 
we essentially perform a contour integral over the function
$\left[A_n - \Inf A_n\right] - \Cuth_n$, which should vanish such that we are
able
to close the contour at infinity. If $\Inf \Cuth$ does 
not vanish, we need to subtract it off, such that the contour integral over
$\left[A_n - \Inf A_n\right] - \big[\Cuth_n - \Inf \Cuth_n \big]$ vanishes,
and Cauchy's theorem can be applied.

\begin{figure}[t]
\centerline{\epsfxsize 2 truein\epsfbox{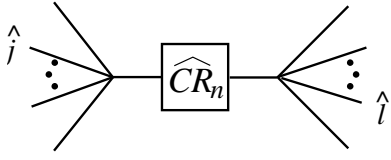}}
\caption{Diagrammatic representation of overlap contributions.
Each overlap diagram corresponds to a physical channel. }
\label{OverlapGenericFigure}
\end{figure}

Double counting between the rational
terms in the cut completion $\Cuth_n$ and the rational remainder
is avoided by the `overlap terms' $\Overlap_n$,
\be
\Overlap_n \equiv
\sum_{{\rm poles}\ \alpha} \Res_{z=z_\alpha}  {\CuthRat_n(z)\over z}  \, ,
\ee
where in general every channel that gets shifted according to
\eqn{SpinorShift} contributes to the sum, as illustrated in
fig.~\ref{OverlapGenericFigure},
but for specific cut completions,
individual channels may vanish. $\Overlap_n$ can be computed
straightforwardly from the cut completion.

\begin{figure}[t]
\centerline{\epsfxsize 5.5 truein\epsfbox{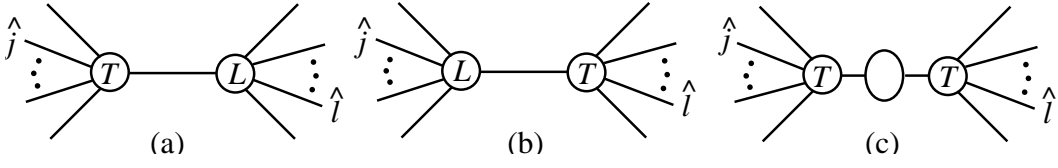}}
\caption{ Schematic representation of one-loop recursive
contributions to \eqn{RationalRecursion}.
The labels `$T$' and `$L$' refer to tree and loop
vertices.}
\label{LoopGenericFigure}
\end{figure}

Finally, the sum over residues of the rational part again results in
a recursion relation, quite analogous to the tree-level case,
\begin{eqnarray}
 - \sum_{{\rm poles}\ \alpha} \Res_{z=z_\alpha} {\Vertex_n(z)\over z}
&\equiv&
R^D_n(k_1,\ldots,k_n) \nonumber\\
 &=& \hskip -.1cm
 \sum_{r,s}\, \sum_{h} \Biggl\{
 A^{\tree \, ,h}_L(z= z_{rs}) {i\over K_{r\ldots s}^2}
\Vertex_R^{-h}(z = z_{rs})
 \nonumber \\
&& \qquad \quad + \,
\Vertex_L^h(z = z_{rs}) {i\over K_{r\ldots s}^2}
A^{\tree \, , -h}_R(z= z_{rs})
\nonumber \\
&& \qquad \quad + \,
A^{\tree \, , h}_L(z= z_{rs}) {i \Fact(K_{r\ldots s}^2) \over K_{r\ldots s}^2}
A^{\tree \, ,-h}_R(z= z_{rs}) \Biggr\}
 \,.  \label{RationalRecursion}
\end{eqnarray}
In the first two terms the scalar propagator connects the rational part of an on-shell
loop amplitude with an on-shell tree amplitude, the last term corresponds to
a one-loop correction to the propagator \cite{Bern:1995ix}. \Eqn{RationalRecursion}
is schematically illustrated in fig.~\ref{LoopGenericFigure}.
However, for a given shift, we may
encounter non-standard channels with as yet unknown factorization behavior,
which arise from two-particle channels with like-helicity gluons,
\be
R^D_n \equiv R^{D, \rec}_n + R^{D, \nonst}_n \, ,
\ee
where $R^{D, \nonst}$ involves
$R_3(k_i^\pm, k_{i+1}^\pm, - \hat{K}_{i\, (i+1)})$ channels ($i$ or $i+1 = \hat{j}$ or
$\hat{l}$ for a $\Shift{j}{l}$-shift)\footnote{$R_3(k_i^\pm, k_{i+1}^\mp, - \hat{K}_{i\, (i+1)}) = 0$,
for either helicity of $\hat{K}_{i\,(i+1)}$, and
$R_3(k_i^-, k_{i+1}^-, -\hat{K}) = 0$ if $i$ or $i+1 = \hat{l}$, as well as
$R_3(k_i^+, k_{i+1}^+, -\hat{K}) = 0$ if $i$ or $i+1 = \hat{j}$.\label{foot}}.

In summary, after attempting to construct a recursion relation for the rational
terms, we have,
\be
A_n(0) = {\color{red} \Inf A_n} + \cg \Biggl[ \Cuth_n(0)  - \Inf \Cuth_n
+ R^{D, \rec}_n + {\color{red} R^{D, \nonst}_n}
+  \Overlap_n \Biggr] \, , \label{OneStepRec}
\ee
which has two unknown contributions, the large-parameter contribution $\Inf A$,
and the contribution from channels with not yet fully understood factorization
behavior, $R^{D, \nonst}$.

\subsection{Pairs of Shifts to the Rescue}

As shown in \cite{Berger:2006ci} by empirical study of
known amplitudes, in general we can find
shifts (\ref{SpinorShift}) which are either free of large-parameter contributions
or free of contributions from non-standard channels, but not both.
However,
by using a pair of shifts in two \emph{independent} complex parameters, we can
construct a recursion relation which determines all unknown terms in
\eqn{OneStepRec}.

We use the primary shift \eqn{SpinorShift}, and an
auxiliary shift of two different legs,
\begin{equation}
\Shift{a}{b}:\hskip 2 cm
\tlambda_a \rightarrow \tlambda_a - w\tlambda_b \,,
\hskip 2 cm
\lambda_b \rightarrow \lambda_b + w\lambda_a \, .
\label{AuxiliarySpinorShift}
\end{equation}
We obtain the following recursion relations for the amplitude,
\bea
A_n(0) & = & {\color{red} \InfPart{\Shift{j}{l}}{A_n}} +
\cg \Biggl[ \Cuth_n(0)  - \InfPart{\Shift{j}{l}} {\Cuth_n}
+ \DiagrammaticRationalSrec{\Shift{j}{l}}_n  +  \Overlap_n^{\Shift{j}{l}}\Biggr] \, ,
\label{primary} \\
A_n(0) & = & \cg \Bigl[ \Cuth_n(0) - \InfPart{\Shift{a}{b}}{\Cuth_n}
                    + \DiagrammaticRationalSrec{\Shift{a}{b}}_n
                    + {\color{red} \DiagrammaticRationalSnonst{\Shift{a}{b}}_n}
                    + \Overlap_n^{\Shift{a}{b}}\Bigr] \, ,
                    \label{secondary}
\eea
where we have indicated with additional superscripts which shift has been employed.
Applying the primary shift \eqn{SpinorShift} to
 the auxiliary recursion (\ref{secondary}), we can extract the large-parameter
 behavior of the \emph{primary shift} according to \eqn{Infdef},
\be
{\color{red} \InfPart{\Shift{j}{l}}{A_n}}  =
\cg \Bigl[ \InfPart{\Shift{j}{l}}{\Cuth_n} -
\InfPart{\Shift{j}{l}}{\bigg(\InfPart{\Shift{a}{b}}{\Cuth_n}\bigg)}
                    + \InfPart{\Shift{j}{l}}{\DiagrammaticRationalSrec{\Shift{a}{b}}_n}
                    + \InfPart{\Shift{j}{l}}{\Overlap_n^{\Shift{a}{b}}}\Bigr] \, ,
\ee
where now all terms on the right-hand side are either known or recursively
constructible, \emph{if}
\be
{ \InfPart{\Shift{j}{l}}{\color{red} \DiagrammaticRationalSnonst{\Shift{a}{b}}_n}}
 = 0 \, . \label{vanish}
\ee
A list of suitable pairs of shifts for which \eqn{vanish} holds,
as well as many examples, can be found in \cite{Berger:2006ci}.

In summary, the bootstrap equation for the full amplitude is given by,
\be
\fbox{ $ \displaystyle
\begin{array}{rcl}
A_n(0) & = & \InfPart{\Shift{j}{l}}{A_n} +
\cg \Biggl[ \Cuth_n(0)  - \InfPart{\Shift{j}{l}} {\Cuth_n}
+ \DiagrammaticRationalSrec{\Shift{j}{l}}_n  +  \Overlap_n^{\Shift{j}{l}}\Biggr]\, ,
\\
& & \\
\InfPart{\Shift{j}{l}}{A_n}  &= &
\cg \Bigl[ \, \InfPart{\Shift{j}{l}}{\Cuth_n} -
\InfPart{\Shift{j}{l}}{\bigg(\InfPart{\Shift{a}{b}}{\Cuth_n}\bigg)}
                    + \InfPart{\Shift{j}{l}}{\DiagrammaticRationalSrec{\Shift{a}{b}}_n}
                    + \InfPart{\Shift{j}{l}}{\Overlap_n^{\Shift{a}{b}}}\Bigr] \, .
\end{array}
$
}
\label{Bootstrap}
\ee
All terms appearing in \eqn{Bootstrap} are either (assumed to be)
known (the cut constructible part),
or can be constructed recursively. Using this approach,
we have computed many previously unknown amplitudes in~\cite{Berger:2006ci}.
We have tested all multiparticle and collinear factorization channels of these
amplitudes. These tests are highly nontrivial, because any omitted or incorrect
terms would spoil the factorization properties.
Furthermore, at six points we have compared our results numerically to the results
of refs.~\cite{Ellis:2006ss,Xiao:2006vt} and found complete agreement.

In summary, the computation of a one-loop amplitude within our bootstrap approach
entails the following steps:
\begin{enumerate}
\item Obtain the pure-cut terms, and complete the cuts such that no spurious
singularities get shifted under the shifts chosen in the next step. It is not
necessary to eliminate all spurious singularities, only those that are not
invariant under the shift(s) and would thus contribute spurious terms to the
recursion.
\item Choose a pair of shifts, a primary shift without non-standard channels
in the recursion, and an auxiliary shift without large-parameter contributions.
Suitable choices of shifts for various helicity configurations are listed
in ref.~\cite{Berger:2006ci}.
\item Compute the large-parameter contributions of the completed cut terms, and
compute the overlap contributions from the primary shift.
\item Compute the direct recursive diagrams arising from the primary shift.
\item Compute the large-parameter contribution from the primary shift
via the auxiliary recursion relation. This includes
the large-parameter contribution under the primary shift of
direct-recursive graphs, cut completed, and overlap terms, computed
via the secondary shift.
\item Add up all contributions according to \eqn{Bootstrap} to obtain
the full amplitude.
\end{enumerate}
We will illustrate these steps with an explicit example in the next section.

\section{A Six-Point Example: $A_{6;1}^{\NeqZero}({-}{-}{-}{+}{+}{+})$}

As an illustrative example, let us compute the six-point amplitude
$A_{6;1}^{\NeqZero}(1^-,2^-,3^-,4^+,5^+,6^+)$~\cite{Berger:2006ci}.
This amplitude has the following flip symmetry,
\begin{equation}
X(1,2,3,4,5,6)\Bigr|_{\rm flip\; 1} \equiv X(3,2,1,6,5,4) \, .
\label{mmmpppflip1def}
\end{equation}

\subsection{Step 1: The Completed-Cut Terms}

We begin with the completed-cut terms, computed in ref.~\cite{Bern:2005hh}.
In general, the terms containing (poly)logarithms will have to be
completed with suitable rational terms, however, in \cite{Bern:2005hh}
this cut-completion has already been performed, yielding,
\bea
   \Cuth_6(1^-,2^-,3^-,4^+,5^+,6^+)  &=&
 \frac{1}{3 \cg}\,A_{6;1}^{\,\NeqOne}(1^-,2^-,3^-,4^+,5^+, 6^+)
\nn\\
&& \hskip0.0cm
 + {2\over 9} A_6^\tree(1^-,2^-,3^-,4^+,5^+, 6^+)
 + \hat{C}_6^a + \hat{C}_6^a \Bigr|_{\rm flip\; 1}
\,,~~ \label{Cuth6}
\eea
where
\bea
  \hat{C}_6^a &=&
{i\over3} \Biggl[ { \spa1.2\! \spa2.3\!\spb2.4 \!\spab1.{(3\!+\!4)}.2
\bigl[ \spaa3.{4}.{2}.1 s_{234}
      - \spaa3.{2}.{(3+4)}.1 s_{34} \bigr]
  \over \spa3.4\spa5.6 \spa6.1\spb2.3 \spab5.{(3\!+\!4)}.2 }
{ \Lzz ( {-s_{234}\over -s_{34}} ) \over s_{34}^3  }
\nonumber\\
&& \hskip0.0cm
+ { \!\spa3.5\!\spb4.5\!\spb5.6\! \spab5.{(1\!+\!2)}.6
    \bigl[ \sand3.{(5-4)}.6 \!s_{345}
              +\sand3.{(4+5)}.6\! s_{34}\bigr] \,
  \over \spa4.5\spb1.2 \spb1.6 \spab5.{(3+4)}.2 }
{ \Lzz ( {-s_{345}\over -s_{34}} ) \over s_{34}^3  } \Biggr]
%
\,. \nonumber\\
&& \hskip0.0cm{~} \label{Cuth6a}
\eea
The first term in \eqn{Cuth6} is proportional to the contribution
of an $\NeqOne$ chiral multiplet in the loop.
This contribution is fully constructible from the four-dimensional
cuts.  The result is~\cite{Bern:2005hh},
\be
A_{6;1}^{\,\NeqOne}(1^-,2^-,3^-,4^+,5^+,6^+) =
S_6^a + S_6^a \Bigr|_{\rm flip\; 1}
\,, \label{AsusyN1mmmppp}
\ee
where
\bea
S_6^a &=&
{i \cg \over 2} \Biggl[
{1\over i} A_6^\tree(1^-,2^-,3^-,4^+,5^+,6^+) \, \Kz( s_{34} )
\label{S6a} \\
&&\hskip0.7cm
- {\sand1.{(2+3)}.4^2\,
\bigl[ \spaa3.{4}.{2}.1 s_{234}
      - \spaa3.{2}.{(3+4)}.1 s_{34}\bigr]
\over \spa{5}.{6} \spa{6}.1 \spb2.3\,s_{234}\,s_{34}
   \sandmm5.{(3+4)}.{2}}
{\Lz ( {-s_{234}\over -s_{34}} ) \over s_{34}  } \nonumber \\
&&\hskip0.7cm
- {\sand3.{(1+2)}.6^2
\bigl[\sand3.{(5-4)}.6 s_{345} + \sand3.{(4+5)}.6 s_{34}\bigr]
\over \spa3.4\spa4.5\spb1.2\spb1.6\,s_{345}\,\sandmm5.{(3+4)}.2 }
{ \Lz ({-s_{345}\over -s_{34}} ) \over s_{34}  } \Biggr]
\,.  \nonumber
\eea
The $\Kz$- and $\Ll_i$-functions can be found in the cited references. Only
$\Ll_2$ contains rational terms.

\subsection{Step 2: Choice of Shift-Pairs}

Following eqs.~(\ref{primary}) and (\ref{secondary}), we choose a $\Shift{1}{2}$
shift as the primary shift,
to avoid non-standard channels (see footnote \ref{foot}).
As the auxiliary shift we choose a $\Shift{3}{4}$ shift,
because from general considerations,
we expect a $\Shift{-}{+}$ shift to be free of large-parameter contributions. The
consistency of our assumptions can be checked after the fact, because,
as already mentioned above, the result has to display the correct factorization
behavior in \emph{all} multiparticle and collinear channels, not just those
channels which have been used to recursively construct the amplitude.

\subsection{Step 3: The Large-Parameter Contribution from Cut Terms and the Overlap Contribution}

It is straightforward to compute the large-parameter contributions
$\Inf \Cuth$ for both shifts from \eqn{Cuth6}. The result is,
\bea
\InfPart{\Shift12}{\Cuth_6(1^-,2^-,3^-,4^+,5^+, 6^+)}
& = &
 {i\over6} { \spa1.2 \spa1.3 \spab3.{(4+5)}.2
  \Bigl[ - \spaa1.2.{(4+5)}.3 + \spa1.3 s_{345} \Bigr]
  \over \spb1.2 \spa3.4 \spa4.5 {\spa6.1}^2
   \, s_{345} \, \spab5.{(3+4)}.2 }
\nonumber\\
&&  +\, {i\over6} { \spa1.2 \spb2.4 \spa1.3
   ( \spa1.2 \spb2.4  - \spa1.3 \spb3.4 )
  \over \spb2.3 \spa5.6 \spa6.1 \, s_{34} \, \spab5.{(3+4)}.2 }
\, ,
\label{Cuth6infpole} \\
\InfPart{\Shift34}{\Cuth_6(1^-,2^-,3^-,4^+,5^+, 6^+)} & = & 0 \, .
\label{cuthpart2}
\eea

\begin{figure}[t]
\centerline{\epsfxsize 6 truein\epsfbox{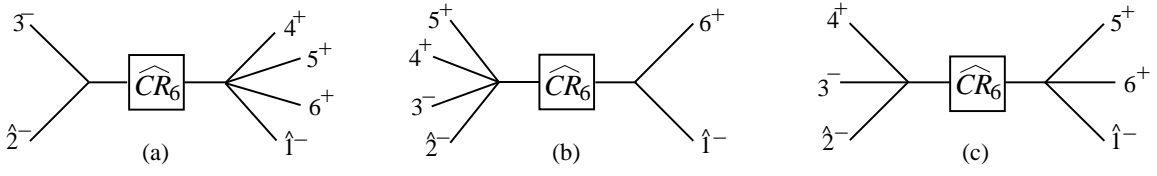}}
\caption{
The overlap diagrams for the $\Shift{1}{2}$ shift.
Diagram (a) vanishes.
}
\label{Overlap3mSixPt12Figure}
\end{figure}

It is also straightforward to compute the overlap contributions for the primary
$\Shift{1}{2}$ shift from \eqn{Cuth6}.
As illustrated in \fig{Overlap3mSixPt12Figure},
there are three channels which can potentially contribute to the
overlap,
\be
\Overlap = \Overlap^{\rm (a)} + \Overlap^{\rm (b)} + \Overlap^{\rm (c)} \,,
\label{O6}
\ee
corresponding to the three
residues of $\CuthRat_6(z)/z$ located
at the following values of $z$,
\be
z^{\rm (a)} = -{\spa2.3 \over \spa1.3}\,, \hskip 2 cm
z^{\rm (b)} = {\spb1.6 \over \spb2.6} \,, \hskip 2 cm
z^{\rm (c)} = -{s_{234} \over \sand1.{(3+4)}.2} \,.
\ee
Evaluating these residues gives us the overlap contributions.
After simplification, they are given by,
\bea
\Overlap^{\rm (a)} &=&
 0
\,, \nn \\
\Overlap^{\rm (b)}
&=&
i \biggl( {1 \over 3\eps} + {8\over9} \biggr)
  { {\spab3.{(1+2)}.6}^3
   \over \spb1.2 \spa3.4 \spa4.5 \spb6.1
   \, s_{345} \, \spab5.{(3+4)}.2 }
\nonumber\\
&&
- {i\over6}\, { \spa1.2 \spab3.{(4+5)}.2 {\spab3.{(1+2)}.6}^2
   \over \spb1.2 \spa3.4 \spa4.5 \, s_{61} s_{345}
   \, \spab5.{(3+4)}.2 }
   \nonumber \\
&& + {i \over 6} \, { \spb4.5 \spb5.6 \spa3.5 \spab5.{(1+2)}.6
\left( \spa3.5 \spb5.6 s_{345} - \spa3.4 \spb4.5 \spab5.{(3+4)}.6
\right) \over \spb1.2 \spa4.5 \spb6.1 \spab5.{(3+4)}.2 } \nn \\
&& \quad \times
{ s_{34} + s_{345} \over s_{34} s_{345} \left( s_{345}-s_{34} \right)^2 }
 - {i \over 6} {\spa1.5 \spb4.5 \spb4.6 \spab1.{(5-6)}.4
\over \spb2.3 \spb3.4 \spa5.6 s_{61} \spab5.{(3+4)}.2 }
 \,, \nn \\
\Overlap^{\rm (c)}
&=&
- i\biggl( {1 \over 3\eps} + {8\over9} \biggr)
 { {\spab1.{(2+3)}.4}^3
    \over \spb2.3 \spb3.4 \spa5.6 \spa6.1
    \, s_{234} \, \spab5.{(3+4)}.2 }
\nonumber\\
&&
+ {i\over6}\, { \spa1.2 \spb2.4  {\spab1.{(2+3)}.4}^2
    \over \spb2.3 \spb3.4 \spa5.6 \spa6.1 \, s_{234} \, \spab5.{(3+4)}.2 }
\nonumber\\
&&
+ {i\over6} \, { \spa1.5 \spa3.4  \spb4.5 \spa1.6  {\spab1.{(2+3)}.4}^2
    \over \spb2.3 \spa5.6 {\spa6.1}^2 \, s_{34} s_{234}
    \, \spab5.{(3+4)}.2 }
 \,.
\label{mmmpppoverlap}
\eea
%

\subsection{Step 4: The Direct Recursive Diagrams}

\begin{figure}[t]
\centerline{\epsfxsize 5.5 truein\epsfbox{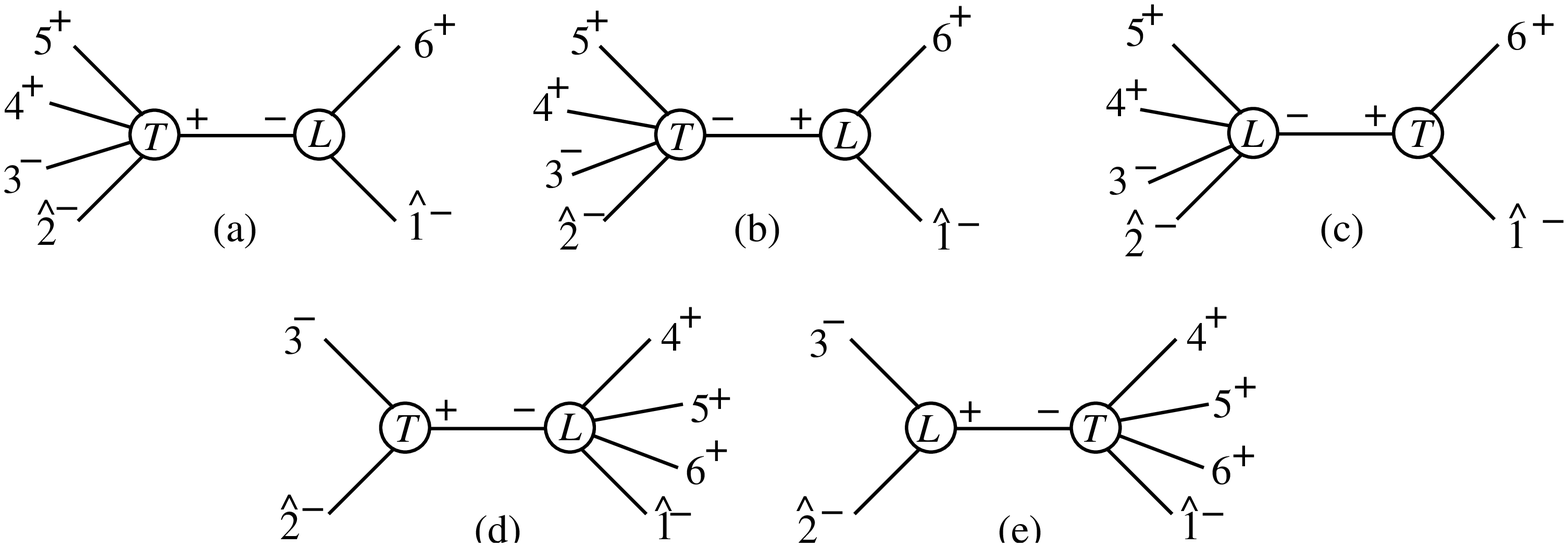}}
\caption{
Some vanishing recursive diagrams for the $\Shift{1}{2}$ shift
of $A^{\NeqZero}_{6;1}(1^-,2^-,3^-,4^+,5^+,6^+)$.
}
\label{Vanish3mSixPt12Figure}
\end{figure}

\begin{figure}[t]
\centerline{\epsfxsize 3.6 truein\epsfbox{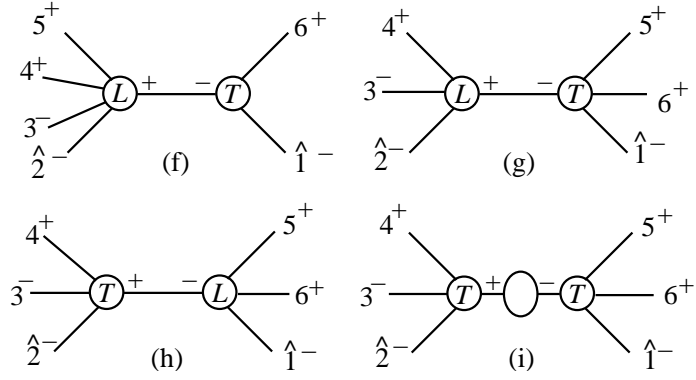}}
\caption{
Non-vanishing recursive diagrams.
Diagram (i) is the factorization-function contribution.
}
\label{Loop3mSixPt12Figure}
\end{figure}

Next we evaluate the recursive diagrams for the $\Shift12$ shift.
The diagrams shown in \fig{Vanish3mSixPt12Figure} vanish,
\be
D_6^{\rm (a)} = D_6^{\rm (b)} = D_6^{\rm (c)}
              = D_6^{\rm (d)} = D_6^{\rm (e)} =  0\,.
\label{Diagram6abcdemmmppp}
\ee
The first two diagrams,
\ref{Vanish3mSixPt12Figure}(a) and \ref{Vanish3mSixPt12Figure}(b),
vanish because the loop vertices
$R_3(6^+, 1^-, \hat K^\pm)$ vanish, as mentioned
in footnote \ref{foot}.
Diagrams \ref{Vanish3mSixPt12Figure}(c) and \ref{Vanish3mSixPt12Figure}(d)
vanish because
\be
A_3^\tree(6^+,\hat 1^-,-\Kh_{61}^+) \propto \spbsh{6}.{\Kh_{61}}^3 = 0\,,
\hskip 1.5 cm
A_3^\tree(\hat 2^-,3^-, -\Kh_{23}^+) \propto \spash{\hat 2}.{3}^3 = 0\,.
\ee
We assume that diagram \ref{Vanish3mSixPt12Figure}(e) vanishes because its
loop vertex is of the same form as the corresponding tree vertex in diagram (d).
For a more extensive discussion of this point we refer to ref.~\cite{Berger:2006ci}.

There are four non-vanishing recursive diagrams,
\be
\DiagrammaticRational_6  = D^{\rm (f)} + D^{\rm (g)}
                         + D^{\rm (h)} + D^{\rm (i)} \,, \label{RD}
\ee
shown in \fig{Loop3mSixPt12Figure}. A list of vertices that enter the recursion
can be found in the appendix of ref.~\cite{Berger:2006ci}. The recursive
diagrams are given by,
\bea
D_6^{\rm (f)} &=&
A_3^\tree(\hat{1}^-,-\hat{K}_{61}^-,6^+) \, {i\over s_{61}}
\, R_5(\hat{2}^-,3^-,4^+,5^+,\hat{K}_{61}^+)
\nonumber\\
&=&
 - i \biggl( {1 \over 3\eps} + {8\over9} \biggr)
  { {\spab3.{(1+2)}.6}^3
   \over \spb1.2 \spa3.4 \spa4.5 \spb6.1
   \, s_{345} \, \spab5.{(3+4)}.2 }
\nonumber\\
&& \null
- {i\over6} \, { s_{345} + s_{34} \over s_{34} s_{345} (s_{345} - s_{34})^2 }
  { \spa3.5 \spb4.5 \spb5.6  \spab5.{(1+2)}.6
   \over \spb1.2 \spa4.5 \spb6.1 \, \spab5.{(3+4)}.2 }
\nonumber\\
&&\hskip1cm \times
   ( \spab3.4.5  \spab5.{(1+2)}.6 + \spab3.5.6 s_{345} )
\nonumber\\
&& \null
+ {i\over3}\, { {\spb4.6}^3 \spab4.{(3+5)}.2
    \over \spb1.2 \spb2.3 \spb3.4 \spa4.5 \spb6.1 \, \spab5.{(3+4)}.2 }
+ {i\over3}\, { {\spb4.6}^2 \spab3.{(1+2)}.6
    \over \spb1.2 \spb3.4 \spa4.5 \spb6.1 \, \spab5.{(3+4)}.2 }
\nonumber\\
&& \null
- {i\over6} \, {\spa3.5  \spb4.5  \spb5.6 \spab3.{(1+2)}.6 \spab5.{(1+2)}.6
      \over \spb1.2\spa4.5 \spb6.1
    \, s_{34} s_{345} \, \spab5.{(3+4)}.2 }
 \,,
\label{Diagram6fmmmppp} \\
D^{\rm (g)} & = &
\, A_4^\tree(\hat{2}^-,3^-,4^+,\hat{K}_{561}^+)
\, {i\over s_{561}}
R_4(\hat{1}^-,-\hat{K}_{561}^-,5^+,6^+)
\nonumber\\
&=&
 \biggl( {1 \over 3\eps} + {8\over9} \biggr)
 A_4^\tree(\hat{2}^-,3^-,4^+,\hat{K}_{561}^+)  \, {i\over s_{561}} \,
 A_4^\tree(\hat{1}^-,-\hat{K}_{561}^-,5^+,6^+)
\nonumber\\
&=&
i \biggl( {1 \over 3\eps} + {8\over9} \biggr)
 { {\spab1.{(2+3)}.4}^3
    \over \spb2.3 \spb3.4 \spa5.6 \spa6.1
    \, s_{234} \, \spab5.{(3+4)}.2  }
\,,
\label{Diagram6gmmmppp} \\
D^{\rm (h)} & = &- D^{\rm (i)} =
  D^{\rm (g)} \,.
\label{Diagram6hmmmppp}
\eea
Diagram~\ref{Loop3mSixPt12Figure}(i) contains the factorization function
contribution~\cite{Bern:1995ix}, which for the scalar loop
case amounts to a vacuum polarization insertion.
The fact that diagrams \ref{Loop3mSixPt12Figure}(g), (h) and (i) are
equal, up to signs, is rather special to this amplitude.

\subsection{Step 5: The Auxiliary Recursion Relation}

\begin{figure}[t]
\centerline{\epsfxsize 5 truein\epsfbox{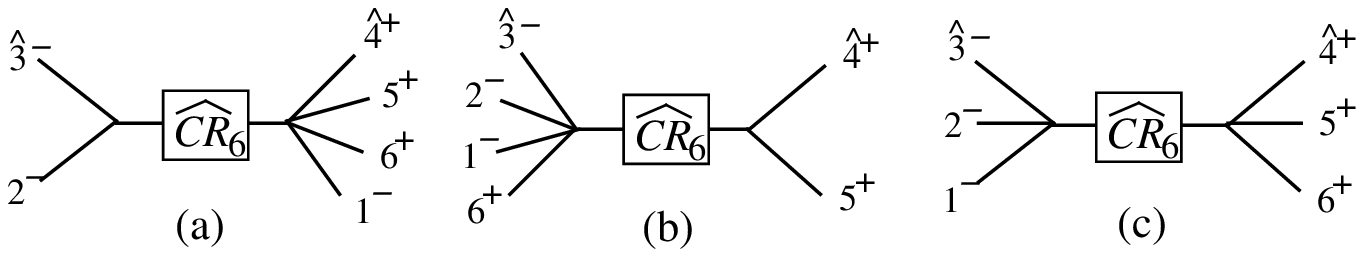}}
\caption{
The overlap diagrams arising from an auxiliary $\Shift34$ shift in
$A^{\NeqZero}_{6;1}(1^-,2^-,3^-,4^+,5^+,6^+)$. }
\label{Overlap3mSixPt34Figure}
\end{figure}

\begin{figure}[t]
\centerline{\epsfxsize 6.5 truein\epsfbox{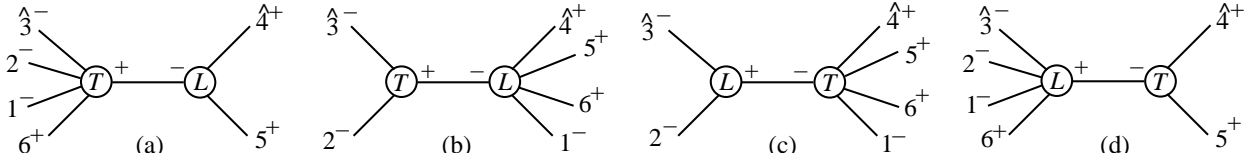}}
\caption{
The recursive diagrams arising from an auxiliary $\Shift34$ shift in
$A^{\NeqZero}_{6;1}(1^-,2^-,3^-,4^+,5^+,6^+)$.
Diagram (c) has non-standard complex singularities.  }
\label{Loop3mSixPt34Figure}
\end{figure}

The diagrams contributing to the overlap and direct recursion from
the auxiliary $\Shift{3}{4}$ shift are shown in figs.~\ref{Overlap3mSixPt34Figure}
and \ref{Loop3mSixPt34Figure}.

It turns out that in this case the large-parameter behavior under the primary
shift of the overlap contribution from the auxiliary shift cancels
against that of the cut-completed part,
\be
\InfPart{\Shift12}{\Cuth_6(1^-,2^-,3^-,4^+,5^+,6^+)} +
 \InfPart{\Shift12}{\Overlap_6^{\Shift34}(1^-,2^-,3^-,4^+,5^+,6^+)} = 0 \, .
 \ee
Furthermore, $\InfPart{\Shift34}{\Cuth_6}$
vanishes according to \eqn{cuthpart2}.
The only remaining contribution
comes from the direct recursive diagrams. This simplification, however, does
not occur in general, and an example where all the aforementioned terms contribute can be
found in ref.~\cite{Berger:2006ci}.

Diagram (c) in fig.~\ref{Loop3mSixPt34Figure}
has non-standard complex singularities. However, according to
\eqn{Bootstrap}, we only need to keep contributions that are
non-vanishing under the primary $\Shift{1}{2}$ shift. It turns out that
here only diagram (d) of fig.~\ref{Loop3mSixPt34Figure}
contributes:
\be
\InfPart{\Shift{1}{2}}{A_{6;1}^{\NeqZero}(1^-,2^-,3^-,4^+,5^+,6^+)}
= \InfPart{\Shift{1}{2}}{ R(1^-,2^-,\hat{3}^-,\hat
       K_{45}^+,6^+)} {i\over s_{45}}
    A_3^\tree(-\hat{K}_{45}^-,\hat{4}^+,5^+)\,,
\label{LargeZRecursion6Pt34}
\ee
where the hatted momenta denote momenta shifted under the auxiliary $\Shift{3}{4}$
shift with $w$ frozen to the value
\be
w^{\rm (d)} = - {\spa4.5 \over \spa3.5} \,.
\label{w34frozen}
\ee
We obtain
\be
\InfPart{\Shift12}{A^{\NeqZero}_{6;1}(1^-,2^-,3^-,4^+,5^+,6^+)}
=
 i\,{\cg\over3}{\spa1.3^3\spa2.6\over\spa1.6^2\spa3.4\spa4.5\spa5.6\spb1.2}\, .
\label{Boundarymmmppp12}
\ee

\subsection{Step 6: Adding Up All Contributions}

Upon adding up all contributions, eqs. (\ref{Cuth6}), (\ref{Cuth6infpole}),
(\ref{O6}), (\ref{RD}), and (\ref{Boundarymmmppp12}),
following \eqn{Bootstrap}, we obtain
the final result for the amplitude,
\be
A^{\NeqZero}_{6;1}(1^-,2^-,3^-,4^+,5^+,6^+) =
\cg \Bigl[  \Cuth_6(1^-,2^-,3^-,4^+,5^+,6^+) + \Remaining_6(1^-,2^-,3^-,4^+,5^+,6^+) \Bigr] \,,
\label{FullA6mmmppp}
\ee
where $\Cuth_6(1^-,2^-,3^-,4^+,5^+,6^+)$ is given in \eqn{Cuth6}, and
\be
\Remaining_6(1^-,2^-,3^-,4^+,5^+,6^+) = \Remaining_6^a  + \Remaining^a_6 \Bigr|_{\rm flip\; 1} \, ,
\label{mmmpppRsimple}
\ee
where
\bea
\Remaining_6^a &=&
 {i\over6} { 1 \over \spb2.3 \spa5.6 \, \spab5.{(3+4)}.2 }
   \Biggl\{   - { {\spb4.6}^3 \spb2.5 \spa5.6 \over \spb1.2 \spb3.4 \spb6.1 }
  - { {\spa1.3}^3 \spa2.5 \spb2.3 \over \spa3.4 \spa4.5 \spa6.1 }
\nonumber\\
&&\hskip1cm \null
 + { {\spab1.{(2+3)}.4}^2 \over \spb3.4 \spa6.1 }
          \biggl( { \spab1.{2}.4 - \spab1.{5}.4 \over s_{234} }
          + { \spa1.3 \over \spa3.4 }
          - { \spb4.6 \over \spb6.1 } \biggl)
\label{mmmpppRasimple}\\
&&\hskip1cm \null
  - { {\spa1.3}^2  ( 3  \spab1.{2}.4 + \spab1.{3}.4 )
     \over \spa3.4 \spa6.1 }
  + { {\spb4.6}^2 ( 3  \spab1.{5}.4 + \spab1.{6}.4 )
     \over \spb3.4 \spb6.1 }
    \Biggr\}
\,. \hskip 1 cm
\nn
\eea
The result~(\ref{mmmpppRsimple}) is manifestly symmetric,
not only under the flip~(\ref{mmmpppflip1def}), but also under
the second flip symmetry, involving spinor conjugation,
\be
X(1,2,3,4,5,6)\Bigr|_{\rm flip\ 2} \equiv X^*(6,5,4,3,2,1).
\label{mmmpppflip2def}
\ee

\section{Summary and Outlook}

Above, we have presented a method for the computation of one-loop
multi-gluon amplitudes that combines information from unitarity, to
determine the cut containing, (poly)logarithmic terms of the amplitude,
and factorization, to obtain the rational remainder via
on-shell recursion relations. Although the construction of on-shell
recursion relations is hindered by several obstacles at the one-loop
level, our bootstrap method systematically bypasses these. The method
is valid for arbitrary helicity configurations, and relies only
on general properties of the gauge theory. We are therefore optimistic
that it can be extended to include fermions, and massive partons.

In certain cases, the on-shell recursion relations can be solved explicitly~\cite{FordeProcs}.
All-multiplicity results have been obtained for several multi-gluon amplitudes,
with all helicities positive or all but one helicity positive~\cite{Bern:2005hs},
for MHV amplitudes~\cite{Berger:2006vq,Forde:2005hh}, and
for split-helicity configurations with three adjacent negative helicities and
the remainder of positive helicity~\cite{Berger:2006ci}.
The study of the properties of these all-multiplicity amplitudes may reveal
some more hidden structure of the underlying gauge theory.

Of course, the aforementioned work is just a starting point. As said above,
it is desirable to extend the method to partons other than gluons, as well
as fully automatize the computations. Furthermore, a better understanding of
complex factorization properties in the non-standard channels would definitely
be useful and might even lead to the applicability of recursion relations to
higher loop order.

\begin{quote}
``One of the most remarkable discoveries in elementary particle physics has been
that of the existence of the complex plane.''
\end{quote}
\vspace*{-6mm}
\hfill from J. Schwinger~\cite{Schwinger} (1970)

\Acknowledgements

I thank Zvi Bern, Lance Dixon, Darren Forde, and David Kosower for
a very fruitful collaboration. I would also like to thank the
conveners of CAQCD06, SUSY06, the Loopfest V, and VLCW06 for inviting me to
very stimulating and interesting meetings.

Due to lack of space and due to the vast amount of literature on this topic,
the list of references below is necessarily incomplete. A more
complete list can be found in \cite{Berger:2006ci,Berger:2006vq}.

\end{document}